\documentclass[%
 reprint,
superscriptaddress,
 amsmath,amssymb,
 aps,
prl,
]{revtex4-1}

\usepackage{soul}
\usepackage{xcolor}
\usepackage{graphicx}
\usepackage{dcolumn}
\usepackage{bm}

\begin{document}

\preprint{APS/123-QED}

\title{Second-Scale Coherence Measured at the Quantum Projection Noise Limit with Hundreds of Molecular Ions}

\author{Yan Zhou}
\thanks{These authors contributed equally.\\Correspondence to: yan.zhou@jila.colorado.edu or\\ yuval.shagam@jila.colorado.edu.}
\author{Yuval Shagam}
\thanks{These authors contributed equally.\\Correspondence to: yan.zhou@jila.colorado.edu or\\ yuval.shagam@jila.colorado.edu.}
\author{William B. Cairncross}
\author{Kia Boon Ng}
\author{Tanya S. Roussy}
\author{Tanner Grogan}
\author{Kevin Boyce}
\author{Antonio Vigil}
\author{Madeline Pettine}
\affiliation{JILA, NIST and University of Colorado, and Department of Physics, University of Colorado, Boulder Colorado 80309-0440, USA}

\author{Tanya Zelevinsky}
\affiliation{Department of Physics, Columbia University, New York, New York 10027-5255, USA}
\author{Jun Ye}

\author{Eric A. Cornell}
\affiliation{JILA, NIST and University of Colorado, and Department of Physics, University of Colorado, Boulder Colorado 80309-0440, USA}

\begin{abstract}
Cold molecules provide an excellent platform for quantum information, cold chemistry, and precision measurement. Certain molecules have enhanced sensitivity to beyond Standard Model physics, such as the electron's electric dipole moment ($e$EDM). 
Molecular ions are easily trappable and are therefore particularly attractive for precision measurements where sensitivity scales with interrogation time. Here, we demonstrate a spin precession measurement with second-scale coherence at the quantum projection noise (QPN) limit with hundreds of trapped molecular ions, chosen for their sensitivity to the $e$EDM rather than their amenability to state control and readout.   Orientation-resolved resonant photodissociation allows us to simultaneously measure two quantum states with opposite $e$EDM sensitivity, reaching the QPN limit and fully exploiting the high count rate and long coherence.
\end{abstract}

\maketitle

Molecular ions are being used in the search for the electron's electric dipole moment ($e$EDM) \cite{Cairncross2017}, with the potential to break the current sensitivity limit \cite{Andreev2018}. The flexibility of ion traps enables  the probing of a coherent superposition  at long times, directly improving sensitivity and reducing the susceptibility to systematic uncertainty. This intrinsic advantage provides the freedom to choose molecules such as HfF$^+$ and ThF$^+$, which have enhanced sensitivity to the $e$EDM, leveraging the long interrogation time to search for physics beyond the Standard Model. Molecular ions can also benefit precision measurements in related experimental investigations, such as electroweak interaction \cite{Quack2008}, fundamental symmetry violation \cite{Flambaum2014,Cairncross2017,Andreev2018,Hinds2011,Flambaum19,Jayich19}, variation of fundamental constants \cite{Hanneke2016,Safronova2017,Kondov2019}, and dark matter searches \cite{VAVRA2014169,Safronova2017,Stadnik2018}. While directly laser coolable molecules \cite{Barry2014,Collopy2018,Anderegg2018,Truppe2017,Kozyryev2017} and assembled molecules \cite{Ni2008,Park372} are typically chosen for their efficient state preparation and readout rather than their measurement utility, for our molecular ion systems, we must separately devise efficient state preparation \textit{and} detection methods to fully exploit their amenability to trapping and make the most precise measurement possible. Specifically, we desire to measure coherent quantum-state superpositions at the fundamental limits set by the state lifetime and the quantum projection noise (QPN) limit. A challenge in precision metrology is that harvesting the QPN-limited sensitivity becomes ever harder with increasing count rate, as technical noise becomes proportionally more significant. We present here a noise-immune scheme for extracting the internal quantum phase in large samples of molecular ions.

\begin{figure}[b]
\includegraphics[trim={0cm 4.5cm 13.5cm 3.5cm}, clip, width=0.7\columnwidth]{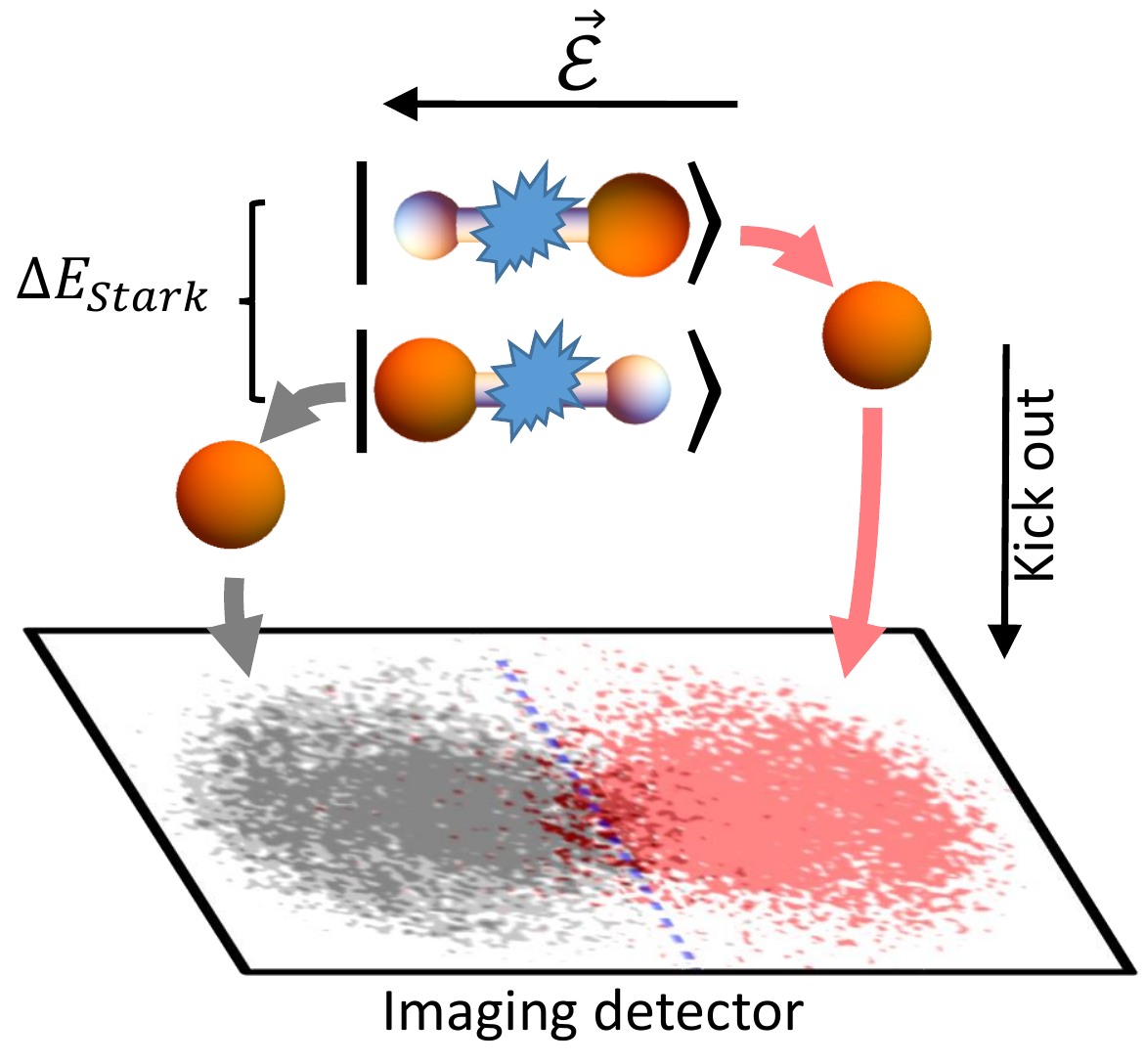}
\caption{Molecules are prepared in two oriented states. Upon dissociation, the photofragments are ejected in the direction of the molecular orientation and kicked out toward an imaging detector. The 2D image of Hf$^+$ ions after HfF$^+$ dissociation is pictured. The quantization axis is defined by the applied electric field $\vec{\varepsilon}$. $\Delta{E}_\mathrm{Stark}$ is the energy difference between the molecular orientations.}
\label{Figure1}
\end{figure}

We employ efficient internal state cooling to significantly enhance the population in the desired science state for the EDM search with both ThF$^+$ and HfF$^+$. To fully utilize the large number of signal ions, we use angularly resolved photodissociation to measure two quantum states of opposite molecular orientation in the same experiment cycle to differentially isolate the coherent superposition signal from technical noise, extracting it near the QPN limit (Fig.\ \ref{Figure1}). For an $e$EDM-like experiment using trapped molecular ions we demonstrate that these improvements decrease our statistical uncertainty by more than an order of magnitude. Our second-scale coherence is now measured near the QPN limit, leading to a statistical sensitivity of {0.3~m$\mathrm{Hz\sqrt{h}}$ with HfF$^+$ (corresponding to an $e$EDM uncertainty of $2.7\times 10^{-29}$ $e\,\mathrm{cm\sqrt{h}}$) as compared to Ref.\ \cite{Cairncross2017} where the best sensitivity was 14.5~m$\mathrm{Hz\sqrt{h}}$.}

\begin{figure*}
\centering
\includegraphics[trim={0cm 0cm 0 0cm}, clip, width=1.3\columnwidth]{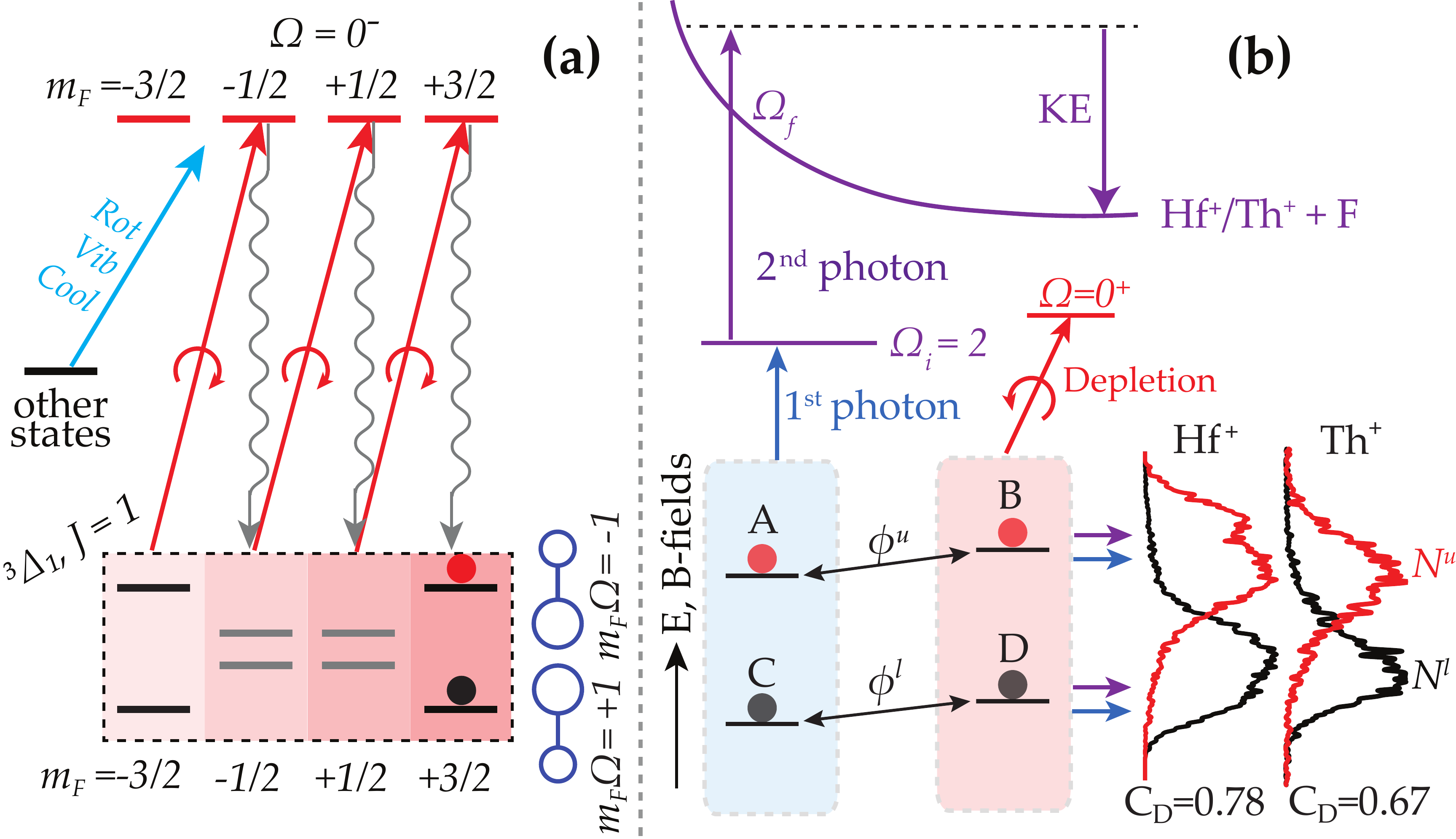}
\caption{Preparation of the initial quantum state and orientation-selective photofragmentation of  HfF$^+$ and ThF$^+$. (a) We use circularly polarized light via an $\Omega=0^-$ electronic state to pump to the fully stretched $m_F$ states of the lowest rovibrational state of $^3\Delta_1$. (b) After the spin-precession sequence, we project the fully stretched m$_F$ states by depletion of the other stretched states. This is done with a circularly polarized excitation to the $\Omega=0^+$  from which the possibility of decay back to $^3\Delta_1$ is small. Additional molecular species specific optical and microwave fields provide rotational and vibrational cooling (see S.I.).  This is followed by two-color photodissociation of the $e$EDM-sensitive, $^3\Delta_1$ states, where the product of $m_F$ and $\Omega$ determines the molecular orientation. The first REMPD photon excites the molecules to the fully stretched bound intermediate state $|\Omega_{i} = 2\rangle$, which maintains their orientation. The second REMPD photon couples the intermediate state to the continuum states resulting in oriented photofragments, Hf$^+$ or Th$^+$ and F atoms, and determines the kinetic energy (KE). The metallic ions from each molecular orientation are individually mapped onto an imaging detector with the resulting 1D distributions. Their orientation contrasts, $C_{D}$, (defined in the Supplemental Material) are 78\% and 67\% respectively.}
\label{Figure3}
\end{figure*}

\textit{State preparation}. The high multiplicity of closely spaced levels in molecules such as HfF$^+$ and ThF$^+$ means that any desired coherent signal is highly diluted at the beginning of an experiment cycle. Careful consideration of the electronic transition branching ratios and leveraging of the ideas developed for optical pumping and laser cooling  of molecules \cite{Stuhl2008,Glockner2015,Staanum2010,Schneider2010,Lien2014} allows us to compensate for significantly off-diagonal Franck-Condon elements such that we can concentrate the population into the  lowest rovibrational quantum state of the $e$EDM-sensitive target $^3\Delta_1$. This increases our signal  while reducing the number of harmful spectator ions, which serve as a possible source of systematic uncertainty and limit the achievable coherence. We prepare the desired spin-polarized stretched states, $|m_{F}=3/2, \Omega=\pm1\rangle$, which are the initial states in our spin precession measurements, by applying circularly polarized light along the $\Omega=0^{-}$ transition [Fig.\ \ref{Figure3}(a)]. This excited state forms an electronically closed system with $^3\Delta_1$ due to parity symmetry. (The quantum number $\Omega$ denotes the projection of the total angular momentum onto the internuclear axis.) Overall, including the rotational and vibrational cooling, we accumulate 60\% of the total population in the spin stretched states for ThF$^+$ (see Supplemental Material \cite{[{See Supplemental Material for state preparation details, which includes Refs. [26-30]}]SIref})\nocite{COSSEL20121,Gresh2016,LOH201249,Loh2013a,Zhou2019} and a comparable fraction for HfF$^+$. We can further prepare a single stretched state with a specific molecular orientation (given by $m_{F}\Omega$), such as $|m_{F}=+3/2, \Omega=+1\rangle$ for diagnostic purposes by depletion of the unwanted state. We label these four relevant fully oriented and fully stretched states $A$, $B$, $C$, and $D$ [Fig.\ \ref{Figure3}(b)]. Additionally, we protect our target science state from contamination by the  remaining spectator ions in various internal molecular quantum states via depletion or repumping of the possible decay channels that may occur over the course of the long interrogation time (see Supplemental Material \cite{SIref}).

\begin{figure*}[]
\centering
\includegraphics[trim={0cm 0.5cm 0cm 0.6cm}, clip,width=1.7\columnwidth]{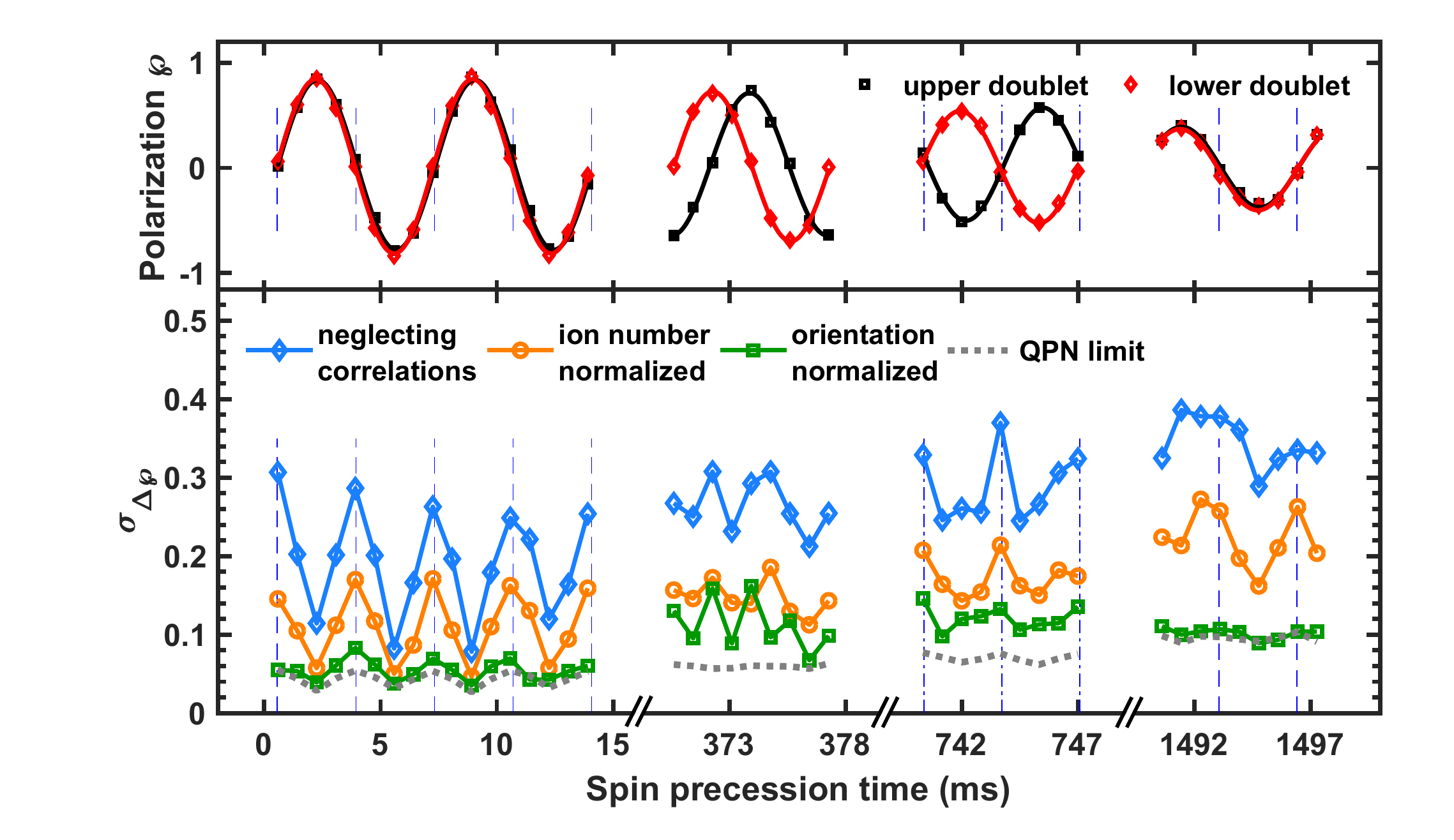}
\caption{Demonstration of a differential measurement at the QPN limit. (Upper) Spin precessions of the upper and lower doublets [Eqs.\ \eqref{eqn:pol1}, and \eqref{eqn:pol2}], which have $\sim0.6$~Hz frequency difference caused by different magnetic moments. (Lower) The scatter in ${\Delta\wp}$. From the scatter of $\wp_u$ and $\wp_l$ determined independently, we would anticipate $\sigma_{\Delta\wp}$ (blue diamonds) to be 5 times above the QPN limit (dotted line). With normalization by each cycle's ion production noise (orange circles), $\sigma_{\Delta\wp}$ is still more than 2 times the QPN limit. By extracting $\Delta\wp$ with simultaneous orientation-selective detection (green squares) we completely eliminate all common-mode noise to bring $\sigma_{\Delta\wp}$ close to the QPN limit when the fringes of the upper and lower doublets are in phase. Spin precession phase noise dominates the out-of-phase measurements ($\sim745$~ms). At early times ($\sim5$~ms), dips in $\sigma_{\Delta\wp}$ are observed at maximum or minimum polarizations [Eq.\ \eqref{eq:qpn}].  Vertical dashed lines mark the zero crossings when the spin precessions are in phase while dash-dotted lines mark them for the out-of-phase case.}
\label{Figure4}
\end{figure*}

With the initial populations prepared in states $A$ and $C$, we initiate the spin precession by applying a $\pi/2$ pulse, lowering the electric bias field as demonstrated in Ref.\ \cite{Cairncross2017}, to create a coherent superposition in both the  upper ($A, B$) and lower ($C, D$) Stark doublets. The same pulse works for both orientations simultaneously. The $e$EDM-like signal contributes to the differential spin precession phase between these doublets. We can compute the phase for the  spin polarizations in the upper and lower Stark doublets by reading out the populations $N_{A}$, $N_{B}$, $N_{C}$, and $N_{D}$ depicted in Fig.\ \ref{Figure3}(b) according to
 \begin{equation}
 {\wp _u} = \frac{{{N_A} - {N_B}}}{{{N_A} + {N_B}}} \sim \mathcal{C} \sin \phi^{u}
 \label{eqn:pol1}
 \end{equation}
 \begin{equation}
 {\wp _l} = \frac{{{N_C} - {N_D}}}{{{N_C} + {N_D}}} \sim \mathcal{C} \sin \phi^{l} 
 \label{eqn:pol2}
 \end{equation}
where $\mathcal{C}$ is the spin precession contrast, and $\phi^{u}$ and $\phi^{l}$ are the spin precession phases of the upper and lower doublets, respectively (Fig. \ref{Figure4}). The coherence time measured in this dataset is $1.9(1)$~s for both molecular orientations and the state lifetime is limited by collisions to 1.14~s.

\textit{Noise-immune state detection}. The most broadly applicable high-yield methods for state-sensitive molecular detection such as resonant photoionization and photodissociation rely on pulsed lasers and are frequently contaminated with noise well in excess of QPN\@. Moreover, they generally allow for the detection of only a single state in a closely spaced multiplet, precluding the possibility of differential measurements. We use resonance-enhanced multiphoton dissociation (REMPD) followed by mass spectrometry to distinguish the Hf$^{+}$ (Th$^{+}$) photodissociation products from background HfF$^+$ (ThF$^+$) \cite{Ni2014,Ng2019}. Laser fluctuations give rise to excess noise on the detected ion number $N$ with standard deviation $\alpha N$, while QPN scales as $N^{1/2}$ \cite{Itano1993}. For instance, our experimental cycle of molecule creation, preparation, and detection involves five pulsed lasers, four of which are frequency doubled. Even with careful monitoring of laser frequencies and intensities, we observe shot-to-shot fractional noise in excess of $\alpha \sim0.2$. For $N\leq25\sim1/\alpha^2$ the QPN limit is reached. For increased $N$, such as our typical sample size $N \gtrsim500$, excess noise from our lasers dominates and the signal-to-noise ratio stops improving \cite{Cairncross2019}.

To circumvent these limitations we developed orientation-selective photofragment imaging, which allows us to count the ion populations in two distinct molecular orientations in a single cycle (Fig.\ \ref{Figure1}). The inherent correlation between these two populations that are prepared and read out by the same laser pulses allows us to make differential measurements at the QPN limit while detecting hundreds of ions in each cycle. Not only are all common-mode fluctuations arising from ionization, state preparation, and detection  canceled by the simultaneous detection of the two molecular orientations, but spin precession phase noise, which may arise from fluctuations in our magnetic bias field, is suppressed as well. Differential measurement is often employed for shot-noise-limited detection in precision metrology \cite{kirilov2013, Lasner2018}. For high detection efficiency of molecules, action spectroscopy and ion detection provide high yields but were not amenable to multiple state detection. Here we implement our new angle-resolved approach to achieve both high detection efficiency and differential noise suppression.

We can extract information about the molecular orientation if the intermediate and final states  inherit the spatial orientation of the ground state, which maps the molecular orientation to momentum-space anisotropy of the photofragments \cite{McDonald2016,Beswick2008, Nandini2010}. Choosing the first REMPD photon such that it excites the molecules to a bound intermediate state with $\Omega_{i} = 2$ maintains the molecular orientation for $M_{i}\neq0$, where $M$ is the projection of total angular momentum on the electric bias field. For the stretched states, $m_{F}$ and $M$ are mapped one to one. We use states with $|M|=|\Omega|$ for the highest orientation contrast between states with opposite orientations (see Supplemental Material \cite{SIref}). The second REMPD photon couples the intermediate state to dissociating states, breaking HfF$^+$ (ThF$^+$) molecules into Hf$^+$ (Th$^+$) ions and neutral F atoms. The spatial distributions of the charged photofragments are shown in Figs.\ \ref{Figure1}, and \ref{Figure3}(b). Experimentally, we obtain similar orientation contrasts for both HfF$^+$ and ThF$^+$.

In a single experimental cycle, we can now simultaneously detect the populations of two states with the same $m_{F}$. In two adjacent cycles, we measure $N_{A}$ and $N_{C}$ followed by $N_{B}$ and $N_{D}$, which we use to compute the spin polarizations according to Eqs.\ \eqref{eqn:pol1} and \eqref{eqn:pol2}. Figure \ref{Figure4} shows experimental measurements of the phase evolution with up to 1.5~s interrogation time. The different magnetic moments of the upper and lower doublet states give rise to a beating between the spin precession fringes.
 
Near zero crossings when the spin precessions of the upper and lower doublets are in phase, $\phi^{u}-\phi^{l} \approx 2n\pi$  (vertical dashed lines in Fig.\ \ref{Figure4}), the phase difference is proportional to the spin polarization difference, $\Delta\wp = \wp_{u} - \wp_{l} \sim \mathcal{C} (\phi^{u} - \phi^{l})$. The $e$EDM sensitivity is ultimately determined by the measurement uncertainty of this spin polarization difference ($\Delta\wp$). If the spin precessions of the upper and lower doublets are evaluated independently, ignoring any correlations, we anticipate the total scatter to be $(\sigma_{\wp_{u}}^2 + \sigma_{\wp_{l}}^2)^{1/2}$, depicted by the blue diamonds in Fig.\ \ref{Figure4}, which is 5 times higher than the QPN limit (dotted gray line) due to excess noise in ion production and detection. Expressed in terms of the spin polarizations [Eqs. \eqref{eqn:pol1}, and \eqref{eqn:pol2}], the QPN limit is 
\begin{equation}
    \label{eq:qpn}
    {\sigma_{\Delta\wp} ^2} = \left( {\frac{{{2}}}{{N_t}}} \right)\left[{{(1+\wp_{u})}\left( {1-\wp_{u}} \right) + {(1+\wp_{l})}\left( {1-\wp_{l}} \right)}\right]
 \end{equation}
where $N_{t}$ is the total number of detected dissociated ions in two adjacent cycles \cite{Itano1993}. {$N_{t}$ is 1400, 1000, 800, and 400, respectively in the four time segments of Fig.\ \ref{Figure4}}. One standard method to reduce excess noise from initial molecular ion production is to normalize for the varying total HfF$^+$ number, which is measured simultaneously with the Hf$^{+}$ number \cite{Shagam2019a}. With ion number normalization (orange circles in Fig.\ \ref{Figure4}), $(\sigma_{\wp_{u}}^2 + \sigma_{\wp_{l}}^2)^{1/2}$ still remains more than 2 times higher than the QPN limit due to excess noise from photodissociation.

Instead of monitoring and correcting for each source of technical noise, we use simultaneous detection of both molecular orientations to normalize all common-mode noise in each experimental cycle. One might expect that a complete normalization of the differential spin polarization $\Delta {\wp}$ requires simultaneous detection of all four states involved. However, simultaneously detecting two states ($A$ and $C$ or $B$ and $D$) is adequate to remove most of the excess noise. Fluctuations in ion production, state preparation, and detection all give rise to excess noise that is positively correlated between $N_A$ and $N_C$ or between $N_B$ and $N_D$. Furthermore, when the two spin precession fringes are in phase, common-mode phase fluctuations contribute to positive correlations between the same pairs of states . When the spin precession fringes are simultaneously near a zero-crossing, we obtain the highest phase sensitivity (vertical dashed lines in Fig.\ \ref{Figure4}). Positive correlations between pairs of these populations cancel excess noise resulting in a measurement of $\Delta\wp$ at the QPN limit, represented by green squares in Fig.\ \ref{Figure4}, both at early and late times ($\sim$1.5~s).

When the spin precessions of the upper and lower doublets are out of phase, the positive correlations from ion production, state preparation, and detection should still cancel to first order at the zero-crossings (vertical dash-dotted lines in Fig.\ \ref{Figure4}). However, common-mode phase noise correlates negatively and cannot be eliminated simultaneously. This phase noise brings the scatter of $\Delta\wp$ significantly higher than the QPN limit, as shown in Fig.\ \ref{Figure4} at $t\approx742$~ms. In our experiment, the early time and late time in-phase measurements are optimal for achieving a QPN-limited $e$EDM sensitivity, cancelling both laser noise and magnetic field noise. The out-of-phase measurements allow us to characterize shot-to-shot phase noise such as arising from magnetic field fluctuations.

In summary, we demonstrate a spin precession measurement at the quantum projection noise limit with hundreds of ions and an interrogation time of 1.5~s. We leverage quantum-state-resolved photochemistry with molecular orientation doublets, a unique demonstration of controlled photofragmentation, to differentially isolate the coherence signal, taking full advantage of the increased count rate of the efficient state preparation scheme. Particularly, for the two quantum states detected by the angle-resolved photodissociation  the electron is polarized with opposite sign, such that the simultaneous differential measurement is sensitive only to the parity odd components such as the $e$EDM. Overall, this scheme has significantly increased the statistical sensitivity of our $e$EDM measurement and may also assist other studies of fundamental symmetries \cite{Safronova2017} or stereochemistry \cite{Heid2019}.

\begin{acknowledgements} We thank J.\ Bohn, L.\ Cheng, R.\ W.\ Field, L.\ Caldwell, R.\ Moszynski and I.\ Majewska for discussions. We thank E.\ Narevicius for loan of equipment. Y.\ S.\ is supported by a postdoctoral fellowship from the National Research Council. W.\ B.\ C.\ acknowledges support from the Natural Sciences and Engineering Research Council of Canada. T.\ Z.\ acknowledges the JILA Visiting Fellows program.  This work was supported by the Marsico Foundation, NIST, and the NSF (Grant No. PHY-1734006).
\end{acknowledgements}

\bibliography{apssamp}

\pagebreak

\begin{center}
\textbf{\large Supplemental Material: Second-Scale Coherence Measured at the Quantum Projection Noise Limit with Hundreds of Molecular Ions}
\end{center}

\subsection{Experimental setup}
Both ThF$^+$ and HfF$^+$ experiments are performed in radio frequency (RF) traps, which are described in Refs.\ \cite{Ng2019,Cairncross2017,Cairncross2019}. The molecular ions are created by resonantly enhanced multiphoton ionization (REMPI) from neutral molecular beams of ThF and HfF, respectively. About 3,000 ThF$^+$ ions or 20,000 HfF$^+$ ions are trapped with ion cloud temperatures of $\sim$10~K. Typical trap frequencies with 50~kHz RF trapping fields are 3~kHz in the $X$,$Y$ directions, and 1~kHz in the $Z$ direction. A rapidly rotating electric field (up to 350~kHz and 60~V/cm) is applied by adding sinusoidal voltages on the trap electrodes. Coils in an anti-Helmholtz configuration provide a linear magnetic field gradient at the center of the trap. An effective rotating magnetic field is generated by coupling ion circular micromotion to the magnetic field gradient \cite{Loh2013a}.

\subsection{Optical pumping in HfF$^+$ and ThF$^+$}
\begin{figure*}[]
\centering
\includegraphics[width=2\columnwidth,trim={0 22cm 10cm 0},clip]{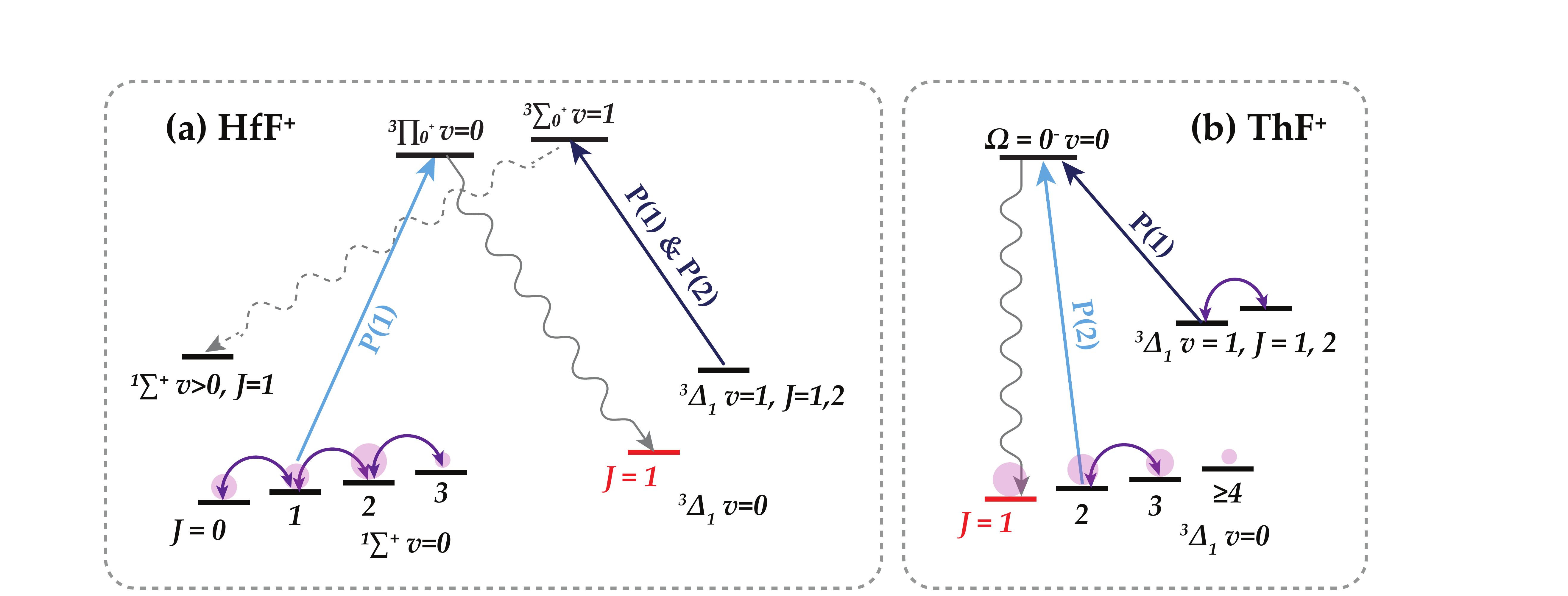}
\caption{Diagram of the transfer process into the $^3\Delta_1$ $v=0$ $J=1$ science state for (a) HfF$^+$ and (b) ThF$^+$. The molecular state population after REMPI is depicted by the purple circles. An $\Omega=0^+$ state ($^3\Pi_{0^+}$) bridges the low-lying $^1\Sigma^+$ and $^3\Delta_1$ states, while an $\Omega=0^{-}$ state couples to $^3\Delta_1$ only. ThF$^+$ ions are created in $^3\Delta_1$, $v=0$ directly, while HfF$^+$ ions are created in $^1\Sigma^+$, $v=0$ and optically pumped to the $^3\Delta_1$ state via the $\Omega=0^{+}$ state. The optical transitions depicted by the cyan arrow combined with microwave transitions indicated by the curved purple lines indicate how we transfer the population into the  $^3\Delta_1$, $v=0$, $J=1$ state while maintaining rotational closure for each molecule. In ThF$^+$, molecules that decay into excited vibrational states ($v=1$) are recovered via the $\Omega=0^{-}$ excited state as shown, while in HfF$^+$ the excited vibrational states are pumped away to $^1\Sigma^+$ via the $^3\Sigma_{0^+}$ state to avoid contamination of $^3\Delta_1$, $v=0$, $J=1$.}
\label{Figure1app}
\end{figure*}

For an extensive discussion of the $\Omega = 0^+$, $\Omega = 0^-$, and other  HfF$^+$ and ThF$^+$ electronic levels, see Refs.\ \cite{COSSEL20121,Ni2014,Gresh2016,Cairncross2019}. Resonance enhanced multiphoton ionization (REMPI) \cite{Zhou2019} creates ThF$^+$ in four rotational states of the ground vibronic state, $^3\Delta_1$, $v=0$ (Fig.\ \ref{Figure1app}b). Counting the hyperfine, Zeeman, and $\Omega$-doublet sublevels, about 100 states are populated. Lessons from laser cooling neutral molecules \cite{Stuhl2008} include: (1) the necessity of $\Delta J=-1$ in all optical transitions to enforce rotational cooling, (2) using fields such as microwaves to mix ground rotational states, which reduces the required number of lasers, and (3) employing vibrational repump lasers to recover population lost to excited vibrational states. While the photon scattering rate of molecules is usually low, especially with the addition of microwave couplings, we enjoy considerable time to perform our cooling because our molecular ions are trapped. Since we are only cooling internal degrees of freedom starting with a few initial rotational states resulting from REMPI, scattering a few photons is sufficient versus $\sim$10$^5$ photons as in most molecular cooling experiments.

\begin{figure}
\centering
\includegraphics[width=1\columnwidth,keepaspectratio]{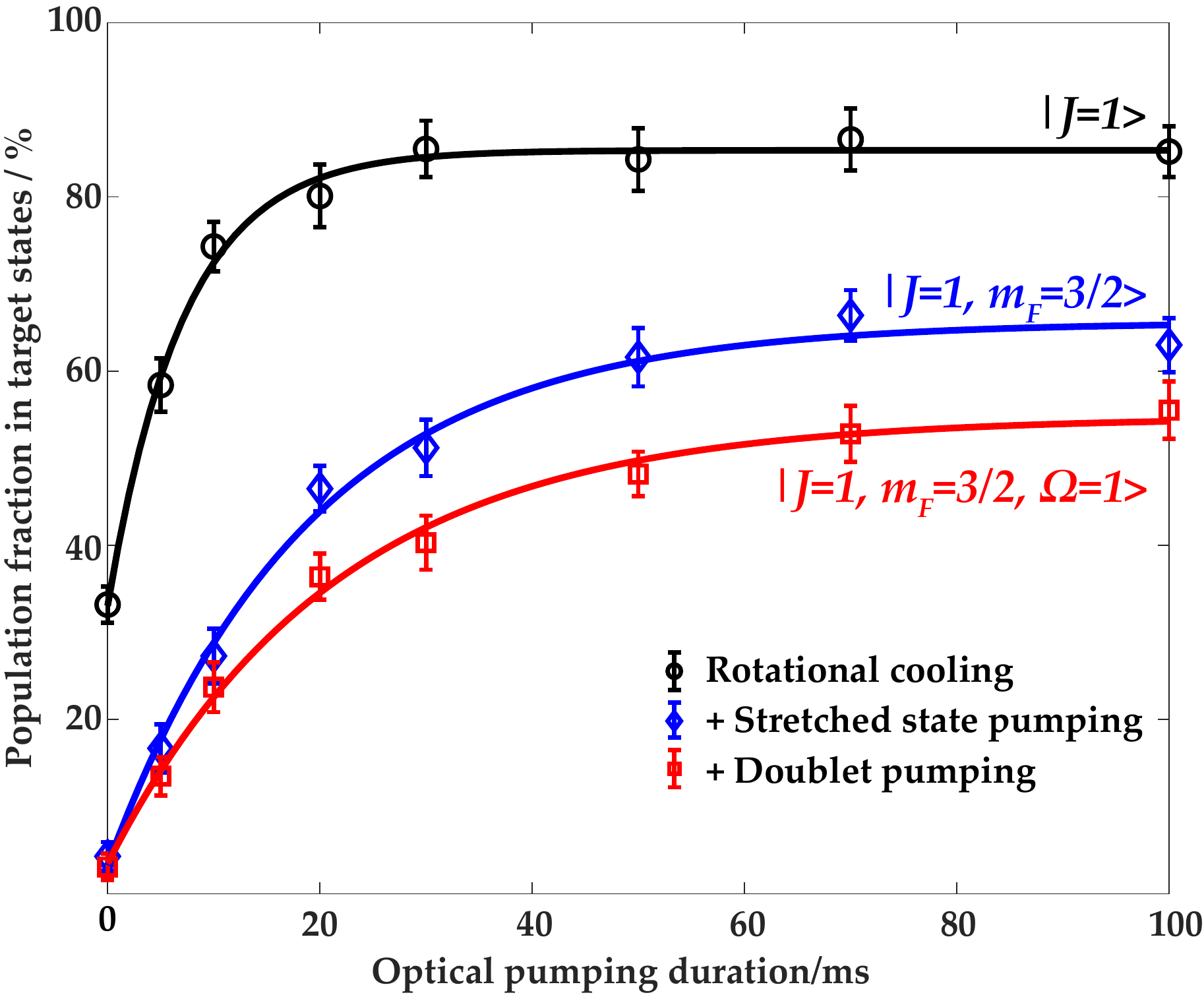}
\caption{Experimental demonstration of successively more elaborate quantum-state preparation of ThF$^+$. Rotational cooling concentrates 85\% of the population in $J=1$ in 20 ms. Including stretched state pumping, we prepare 60\% of total ions in the two stretched states with $m_{F}=3/2$ in 40 ms. Including doublet pumping, we move 50\% of total population into one stretched state with a specific molecular orientation in 50 ms. The error bars are standard errors. }
\label{Figure2app}
\end{figure}

In ThF$^+$ we drive the $J=2 \rightarrow J'=1$ transition to enforce rotational closure, and use microwave couplings to cool molecules in higher rotational states (Fig.\ \ref{Figure1app}b). We transfer 85\% of the population to $^3\Delta_1$, $v=0$, $J=1$ state after 20 ms with additional $v=1$ vibrational repumping (Fig.\ \ref{Figure2app}).

In HfF$^+$ our REMPI scheme \cite{LOH201249,Cairncross2017} initially mostly populates the four lowest rotational states of $^1\Sigma^+$, $v=0$ as shown in Fig.\ \ref{Figure1app}a. We optically pump through $^3\Pi_{0^+}$ which decays mostly towards $^3\Delta_1$ state with microwave coupling to mix $J=0-3$ to transfer the population. Only decay into $J=1$ rotational states of $^3\Delta_1$ is allowed by using the $J=1\rightarrow0$ [P(1)] line, but decay to excited vibrational states is possible. We introduce vibrational `cleanups' on  $^3\Delta_1$, $v=1$, $J=1,2$ states to keep the molecules that have reached excited vibrational states of $^3\Delta_1$ during the transfer process from slowly cascading down the vibrational ladder and contaminating  our science state during the long spin precession experiment.

For both HfF$^+$ and ThF$^+$, we prepare spin-polarized stretched states, $|m_{F}=3/2, \Omega=\pm1\rangle$, via the application of circularly polarized light that is resonant with an $\Omega=0^-$ state. In ThF$^+$ we accumulate 60\% of the total population in the stretched states after 40 ms (Fig.\ \ref{Figure2app}). We can further prepare a single stretched state with a specific molecular orientation ($m_{F}\Omega$) for characterization of the orientation contrast, by applying a microwave field to resonantly depopulate the unwanted stretched state to $^3\Delta_{1}$, $J=2$, where it is immediately recycled via the optical transition to $\Omega=0^{-}$. Overall, we transfer 50\% of trapped molecules to an oriented state after 50 ms (Fig.\ \ref{Figure2app}). In HfF$^+$ the two stretched states are spectrally resolved for optical wavelengths  due to  the larger Stark shift than in the ThF$^+$ experiment and therefore can be directly depleted to produce individual molecular orientations for orientation contrast characterization.

\subsection{Orientation contrast}

The orientation contrast $C_D$ (Fig. 2b of the main text) describes the how well the two molecular orientations are resolved by our resonant photodissociation. We characterized it by leaving a single molecular orientation state  populated after depletion of the other orientation population. We define the orientation contrast  by
\begin{equation}
  {C_D} =\frac{1}{2}\left(\frac{N^u_L-N^l_L}{N^u_L+N^l_L}+\frac{N^l_R-N^u_R}{N^u_R+N^l_R}\right)
 \label{eqn:contrast2}
\end{equation}
where $N^x_y$ is the number of ions found in the anisotropic photofragmentation on the $y$ ($L$eft or $R$ight) side of the detector after preparing the $x$ ($u$pper or $l$ower) doublet.  For the data in this paper we have tallied ion counts either as $L$eft or $R$ight based on a line down the center of the image. With a modest decrease in total count rate, we can improve contrast by designating a finite-width swatch in which ions are assigned to neither category.

{Ultimately the optimal angle-resolved dissociation contrast is achieved for intermediate states with $\Omega=2$ that are fully stretched. The wave functions of the intermediate states are proportional to $D_{2,m_F}^2$ where $D^J_{\Omega,m_F}$ is the Wigner-D matrix. These wave functions are the most acutely oriented for $|m_F| = \Omega=2$. We populate these fully stretched states by using circularly polarized light for the first dissociation photon. The second photon couples the intermediate state to the continuum causing dissociation through multiple channels. We optimized its polarization empirically to maximize $C_D$. For HfF$^+$, we use a pulsed dye laser tuned to 368.3~nm at $\sim1$~mJ/cm$^2$ for the first photon and a 266~nm quadrupled pulsed Nd:YAG laser at $\sim10$~mJ/cm$^2$ for the second photon with the opposite circular polarization relative to the first photon. For ThF$^+$, the first photon is tuned to 307.7~nm at $\sim0.3$~mJ/cm$^2$ and the second photon wavelength is 355~nm at $\sim10$~mJ/cm$^2$.} 

{ After photodissociation, we turn off the trapping fields and pulse on a uniform electric field to accelerate the ions towards the imaging microchannel plate (MCP) detector. We time  the photodissociation lasers such that the bias electric field that orients the molecules is pointing perpendicular to the kick out direction. The charged photofragments are separated in time of flight from the molecules. They are also separated spatially on the detector according to their velocity component that is orthogonal to the kick out direction. The ions acquire the majority of this velocity component from the photodissociation step that depends on the molecular orientation as shown in Fig.\ 1 of the main text. Thus we spatially map the molecular orientation onto the imaging detector.
}

\subsection{Counting dissociated ions on gated imaging MCP with total ion number correction}
To reduce excess noise from ion detection we count the dissociated ions with an imaging MCP detector. By dispersing the dissociated ions onto the 40 mm imaging MCP detector, $>1000$ ions can be counted with high fidelity, with the spatial positions determined as well. In addition, we developed a method to extract the signal of the non-dissociated ions in the same cycle \cite{Shagam2019a}. Thus, we can use the total ion count to suppress the excess noise of ion creation. However, this method cannot reduce the technical noise of the photodissociation. 

\subsection{Fractional technical noise in ion creation and detection}

Both HfF$^+$ and ThF$^+$ ions are created in a two-step process. Firstly, hafnium or thorium plasma is generated by ablating the corresponding solid metal with a 5 ns, several mJ, 532 nm laser pulse from a Q-switched Nd:YAG laser. The hot plasma chemically reacts with SF$_{6}$ in buffer gas to create neutral HfF or ThF molecules in a supersonic beam. There is $\sim$20\% shot-to-shot fluctuation of the molecular beam intensity, primarily originating from non-repeatable metal surface condition of the ablated 

and the fluctuation of the ablation pulse energy. Secondly, the neutral molecules in the beam are ionized by a two-photon resonantly enhanced multiphoton ionization (REMPI) process. Two tunable dye lasers pumped by one Nd:YAG laser generate these two photons. The REMPI adds an additional $\sim$20\% shot-to-shot fluctuation. In all, the number of trapped HfF$^+$ has $\sim$30\% excess noise in our typical experiments. Similar to REMPI, two-photon resonance enhanced multiphoton dissociation (REMPD) for signal readout also requires tunable dye lasers. A $\sim$20\% shot-to-shot fluctuation is observed due to fluctuating pulse energies and laser frequency of our dissociation lasers. 

The proportionality of the excess noise to the signal magnitude can be understood in terms of the efficiency of the REMPD dissociation process. The ion yield from a process such as dissociation is $N=\epsilon N_{molecule}$, exhibiting very high efficiencies in our experiment with $\epsilon$ on the order of 10-50\% of the molecules in the resonant state depending on the specific transitions chosen. However, since $\epsilon<100\%$,  {the number of dissociated molecules} is sensitive to any variation in the dissociation process such as laser fluctuations, creating excess noise such that ${\sigma_N}^2=\epsilon N_{molecule}+\sigma_\epsilon^2 N_{molecule}^2=N+\alpha^2 N^2$ where the first term is shot noise. For large $N$ our signal-to-noise ratio tends toward a constant $N/\sigma_N\rightarrow 1/\alpha$, which we have empirically found to be $\sim$5 for our photodissociation process.

\end{document}